\documentclass{rmf-d}
\usepackage{nopageno,rmfbib,multicol,times,epsf,amsmath,amssymb,cite}
\usepackage[T1]{fontenc} 
\usepackage[]{caption2}
\usepackage{graphicx}
\usepackage{blindtext}
\usepackage{color}
\usepackage[hidelinks]{hyperref}
\usepackage{listings}
\usepackage{nicefrac}
\usepackage{xfrac}
\usepackage{multirow,multicol} 
\usepackage{array} 
\usepackage{float} 
\usepackage{amsmath} 
\def\rmfcornisa{Education in Physics \hfill\rmf\ {\bf E} (*?*) ???--???
\hfill JULY-DECEMBER2025}

\def\rmfcintilla{{\it Rev.\ Mex.\ Fis.\/} {\bf E} (*?*) (????) ???--???}
\clearpage \rmfcaptionstyle 
\begin{document}
\markboth{ RMF EDITORIAL TEAM}{ A \LaTeX template for the RMF, RMF-E, SRMF }

%
%
\title{Adding an Einsteinian motivation to key discussions in an electromagnetism course
\vspace{-6pt}}
\author{L. E. Fuentes-Cobas$^a$ and M. E. Fuentes-Montero$^{b,*}$
}
\address{   $^a$Centro de Investigación en Materiales Avanzados, Chihuahua, México \\
$^b$Facultad de Ciencias Químicas, Universidad Autónoma de Chihuahua, Chihuahua, México\\ $^*$email: mfuentes@uach.mx   }
%
%
\author{ }
\address{ }
\author{ }
\address{ }
\author{ }
\address{ }
\author{ }
\address{ }
\author{ }
\address{ }
\maketitle
%
%
\recibido{23 May 2024}{4 
Nov 2024
\vspace{-12pt}}
\begin{abstract}
\vspace{1em} 
%
%
This paper aims to provide physics teachers with tools to help deepen the understanding of the laws of electromagnetism. The
fundamental contributions of our proposal are: a) to use quotes from mythical characters in the history of science as a motivating
educational resource; b) to promote the discussion of striking and fundamental topics; c) to mention diverse approaches and stimulate
the search for correct answers to provocative questions. Citations from Einstein refer to principal contributions made by Newton,
Maxwell and himself. Emphasis is placed on the cognitive value of differential (local, infinitesimal) analysis of fundamental concepts
(field structure, causality, field relativistic transformations). The unity of electromagnetism is analyzed from the point of view of
special relativity. It is clarified that descriptions suggesting that the magnetic field is dispensable are contrary to the Einstenian
approach: they assume that, to describe the interaction between moving charges, there is a preferred coordinate system for each
particular problem. An introductory presentation of the tensor form of Maxwell’s equations is provided.
 
\vspace{1em}
\end{abstract}
\keys{{\textit{Physics education; Electromagnetic field; Special relativity
}} \vspace{8pt}} 
\begin{multicols}{2}

\section{Introduction}

The nature of the electromagnetic field is one of the topics
that attracts the most students and teachers. The charm of the
subject is seasoned with a touch of indecipherability due to its
intangible nature, and the need of relatively advanced
mathematical tools for its satisfactory assimilation. The
traditional development of electromagnetism courses, based
on teacher presentations while the students take notes, does
little to facilitate the understanding of the content \cite{1,2}. This
paper provides physics teachers with resources to achieve a
deeper knowledge of the laws of electromagnetism. The
fundamental contributions of our proposal are: a) to take
advantage of moments and quotes from mythical characters
in the history of science as a motivating educational resource;
b) to promote the discussion of striking and fundamental
topics; c) to mention diverse approaches and stimulate the
search for correct answers to the problems posed. The
application of the aforementioned teaching resources should
help to fulfil the purpose of the article. These tools have been
identified as important both in classical works \cite{3,4,5} and in
recent programmatic documents of the European Physics
Society \cite{6} and the American Institute of Physics \cite{7}.

The present article considers the electromagnetic field’s
structure, within the framework of classical physics and
special relativity.

The paper has been built from considerations by A.
Einstein. It contains a selection of quotes from his book
"Essays In Science" (EIS) \cite{8}. We start with segments of the
book in which the author of the Theory of Relativity
comments about the contributions of Newton and Maxwell.
Afterward, we develop our suggestion for the application of
Einstein's considerations. The article plan covers the
following topics:

- Stationary fields (Electrostatics and Magnetostatics).

- Time-varying fields and Electromagnetic Induction.

- Relativity of $\mathbf{E}$ and $\mathbf{B}$.

- Maxwell's equations in tensor form.\\

The consistency of unifying electric and magnetic
components into the concept of electromagnetic field within
Einstein's Relativity is reinforced. The educational
importance of historical quotes and the cognitive value of
detailed differential (local, infinitesimal) treatment of
fundamental concepts (field structure, causality, field
relativistic transformations) are highlighted.

\subsection*{Einstein about Newton and Maxwell.}

We reproduce a quotation from the section on Newton in EIS
(\cite{8} pp. 28 and following). Regarding Newton's work on
Kepler's Laws, Einstein expresses his assessment of the
scientific contribution of the author of \textit{Principia}. \\
\noindent
\textit{"Newton's object was to answer the question: Is there such a
thing as a simple rule by which one can calculate the
movements of the heavenly bodies in our planetary system
completely, when the state of motion of all these bodies at one
moment is known? Kepler's empirical laws of planetary
movement, deduced from Tycho Brahe's observations,
confronted him, and demanded explanation. (…) The most
important point, however, is this: these laws are concerned
with the movement as a whole, and not with the question how
the state of motion of a system gives rise to that which
immediately follows it in time; they are, as we should say now,
integral and not differential laws.
The differential law is the only form which completely
satisfies the modern physicist's demand for causality. The
clear conception of the differential law is one of Newton's
greatest intellectual achievements."} \\

The independent variable in Newton's Second Law, $d{\bf{p}} = {\bf{F}} dt$, is time. Einstein highlights the concept of causality as
the characterization of movement generation at the level of
infinitesimal increments of time.

We now transcribe a segment of Einstein's analysis of
Maxwell's contribution to knowledge. This consideration is
related to the first applications of partial differential equations
to study deformable bodies and the supposed waves of ether
in light (EIS, pp. 41 and following).
\newline
\textit{“Neglecting the important individual results which Clerk
Maxwell life-work produced in important departments of
physics, and concentrating on the changes wrought by him in
our conception of the nature of physical reality, we may say
this: Before Clerk Maxwell people conceived of physical
reality— in so far as it is supposed to represent events in
nature—as material points, whose changes consist
exclusively of motions, which are subject to partial
differential equations. After Maxwell they conceived physical
reality as represented by continuous fields, not mechanically
explicable, which are subject to partial differential equations.
This change in the conception of reality is the most profound
and fruitful one that has come to physics since Newton…"} \\

\noindent
In Maxwell's equations, the independent variables are spatial
coordinates and time. The dependent ones are the $\mathbf{E}$ and $\mathbf{B}$
fields. In the electro- and magnetostatic cases ($\nicefrac{\partial}{\partial t}=0$), Maxwell's equations characterize the spatial structure of the
fields. In the general electrodynamic case ( $\nicefrac{\partial}{\partial t}\neq0$),
Maxwell's equations describe the structure and causality of
the electromagnetic field at the infinitesimal level \cite{9}.

 \subsection*{Bohr on causality}

 Causality in electromagnetism, in particular the relationship
of Maxwellian electromagnetism to Newtonian mechanics,
has been referred to by Niels Bohr. In his article on causality
and complementarity \cite{10}, he wrote: \textit{“In classical mechanics,
the forces between bodies were assumed to depend simply on
the instantaneous positions and velocities; but the discovery
of the retardation of electromagnetic effects made it necessary
to consider force fields as an essential part of a physical
system, and to include in the description of the state of the
system at a given time the specification of these fields at every point of space. Yet, as is well known, the establishment of the
differential equations connecting the rate of variation of
electromagnetic intensities in space and time has made
possible a description of electromagnetic phenomena in
complete analogy to causal analysis in mechanics.”} \\

Expressing physical relationships at the level of
differentials of time, space, electronic density,
electromagnetic field, among others, is to reach the level of
detail desired by the current researcher in dynamic systems,
fluids, electromagnetism, quantum physics, etc. One purpose
of this article is to highlight the importance of the differential
approach. \\

In the sections that follow, based on the above Einstenian
quotes (with Bohr’s final touch) we suggest a path for
discussing the electromagnetic field concept, in vacuum, that
tracks the Newton → Maxwell → Einstein's development of
ideas.

\section{The structure of electro- and
magnetostatic fields}

Table \ref{table:1} summarizes the Maxwell equations for time-invariant
electromagnetic fields. \\

The difference between the differential and integral forms
of Maxwell's equations is that the former describes what
happens to the fields in the vicinity of a given point, and the
latter refers to what happens in a finite region of space. The
global structure of the fields (integral equations) is generated
by infinitesimal concatenations characterized by the
differential equations. Newton, Maxwell and Einstein gave
priority to differential equations. In the electromagnetic
theory, the divergence and curl (rotational) differential
operations are of significant interest. The divergence
describes the local variations of a vector field along the
direction in which it points. The curl characterizes the spatial
variations of the field associated with displacements
perpendicular to the vector. The qualitative interpretation of
the divergence and rotational operators in the Purcell-Morin
textbook \cite{11} is illustrative.

Studying electrostatics and magnetostatics in parallel has
methodological advantages, as discussed in the text \cite{12}.
Maxwell equations for electrostatics and magnetostatics are
"inter-crossed". The flux of $\mathbf{B}$ is always zero, and the enclosed
charge determines that of $\mathbf{E}$. The circulation of electrostatic $\mathbf{E}$
is always null, while for $\mathbf{B}$, it depends on the enclosed current. \\

\begin{table*}
\centering
\caption{Maxwell equations for stationary conditions ($\nicefrac{\partial\mathbf{E}}{\partial t}=\nicefrac{\partial\mathbf{B}}{\partial t}=0$)}
\label{table:1}
\vspace{0.3cm}
\begin{tabular}{m{3cm}m{3cm}m{3cm}m{3cm}m{3cm}}
\hline
Case → & \multicolumn{2}{c}{Electrostatics} & \multicolumn{2}{c}{Magnetostatics} \\
\cline{2-5}
  Equation ↓ & Differential  & Integral & Diferential  & Integral \\
  \hline
  a) Gauss's laws for $\mathbf{E}$ and $\mathbf{B}$ & $\nabla\cdot {\bf{E}}=\nicefrac{\rho}{\varepsilon_0}$ & $\oint\mathbf{E}\cdot d\mathbf{a}=\nicefrac{q}{\varepsilon_0}$ & $\nabla\cdot\mathbf{B}=0$ & $\oint\mathbf{B}\cdot d\mathbf{a}=0$ \\
  \hline
 b) Static $\mathbf{E}$ is conservative. Ampere's law for $\mathbf{B}$  & $\nabla\times {\bf{E}}=0$ & $\oint\mathbf{E}\cdot d\mathbf{l}=0$ & $\nabla\times\mathbf{B}=\mu_0\mathbf{J}$ & $\oint\mathbf{B}\cdot d\mathbf{l}=\mu_0I$ \\
 \hline
\end{tabular}
\end{table*} 

The differences between these equations lead to different
solutions and to qualitatively different fields. The general
solution of Maxwell equations for the electrostatic case is
Coulomb's law, expressed in terms of $\mathbf{E}$. For the field
associated with a differential charge \textit{dq}, which is part of a
continuous object, this law is written as in equation (\ref{eqn:1}):

\begin{equation}
\label{eqn:1}
    d\mathbf{E}=\frac{dq\mathbf{\widehat{r}}}{4\pi\varepsilon_0r^2}
\end{equation} 

$\mathbf{E}$ is a polar vector whose lines of force are born in (emerge from) the positive charges, and die in the negative ones. In the magnetostatic case, the solution of the equations for the differential of $\mathbf{B}$, corresponding to a current element
\textit{Id\textbf{l}}, is given by the Biot-Savart law, equation (\ref{eqn:2}):

\begin{equation}
\label{eqn:2}
    d\mathbf{B}=\frac{\mu_0Id\mathbf{l}\times\mathbf{\widehat{r}}}{4\pi r^2}
\end{equation}

$\mathbf{B}$ is an axial vector (or "pseudovector") whose lines of force revolve around the currents, following the well-known "right-hand rule", without being born (or dying) anywhere. The axial nature of $\mathbf{B}$ is expressed by the "$\pmb{\times}$" operator in the Ampere law (Table \ref{table:1}) and in equation (\ref{eqn:2}). The behaviours of
$\mathbf{E}$ and $\mathbf{B}$ under those symmetry transformations including the
inversion, are different. This topic is discussed, for example,
in \cite{13}.

\begin{figure*}[htbp]
\centering
\includegraphics[width=\linewidth]{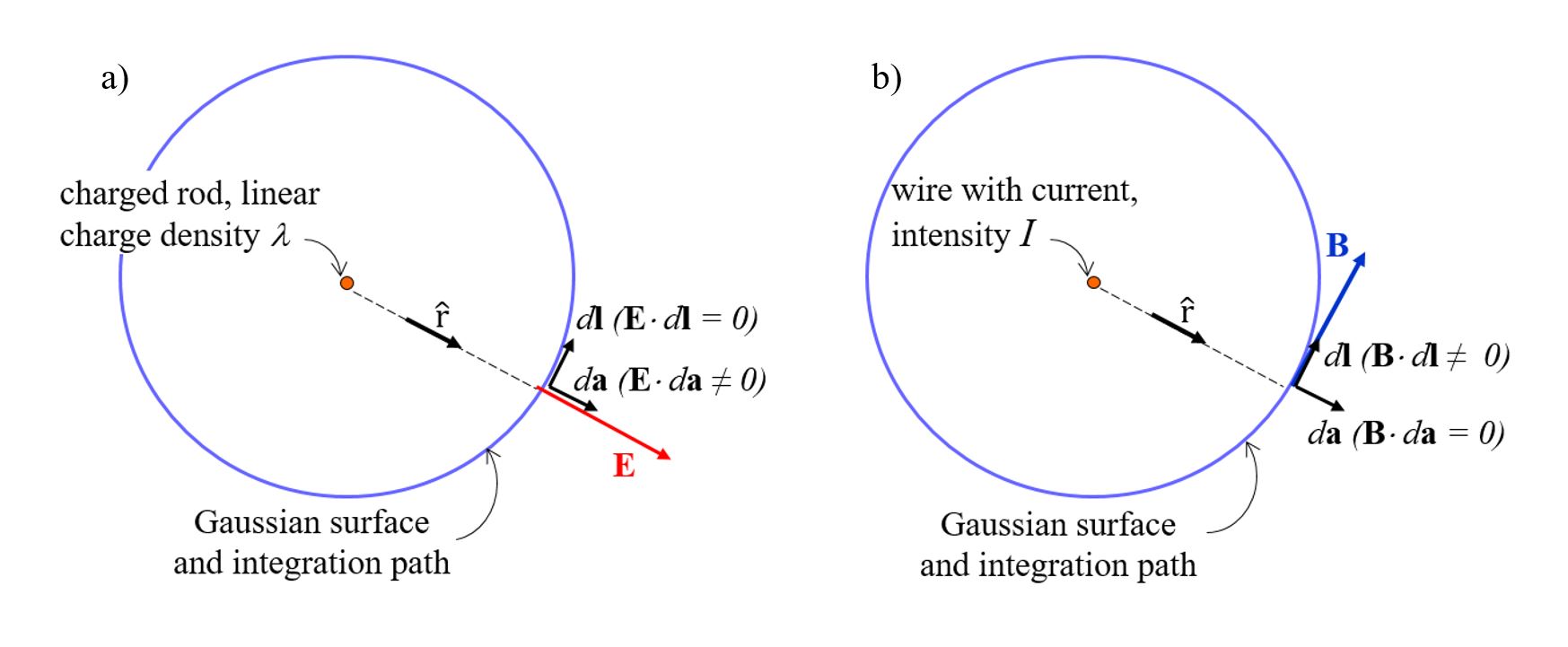}
\caption{Electrostatic (a) and magnetostatic (b) fields associated to infinite rods, of charge and current. The rods come out
perpendicularly to the drawing. Linear $\lambda$ and volumetric $\rho$ charge densities are related by $\lambda =\rho\pi R^2$ . The intensity $I$ and current density $J$ are related by $I=J\pi R^2$. R is the radius of the rods (one insulating, the other conducting). Unit vectors ${\bf{\hat{r}}}$ point from the charge or current elements to the observation points. The (dark blue) circles represent, for Gauss's laws, the intersections of cylindrical Gaussian surfaces with the diagram. For line integrals, these circles are integration paths.}
\label{fig:img1}
\end{figure*}

Figure 1 schematically compares the electrostatic and magnetostatic cases with simple cylindrical symmetries. We use these examples to discuss some ideas. Maxwell equations, coupled with symmetry considerations, lead to the solutions of the $\mathbf{E}$ and $\mathbf{B}$ fields of Figure \ref{fig:img1}. For all space, it follows that E is perpendicular to the charged rod and $\mathbf{B}$ is circular about the current. The modules of these vectors are given by equations (\ref{eqn:3}) and (\ref{eqn:4}).

\begin{equation}
\label{eqn:3}
    E(r)= \begin{Bmatrix}
\frac{\rho r}{2\varepsilon_0} & r < R \\
\frac{\lambda}{2\pi \varepsilon_0 r} & r> R
\end{Bmatrix}
\end{equation}

\begin{equation}
\label{eqn:4}
    B(r)= \begin{Bmatrix}
\frac{\mu_0 Jr}{2\varepsilon_0} & r < R \\
\frac{\mu_0 I}{2\pi r} & r> R
\end{Bmatrix}
\end{equation}

In free space, the geometries of the $\mathbf{E}$ and $\mathbf{B}$ fields are different (“radial” versus “circular”), but both are irrotational ($\nabla\times\mathbf{E}=\nabla\times\mathbf{B}=0$. For example, the vector $\mathbf{E}$ satisfies the condition

\begin{equation}
\label{eqn:5}
    \frac{\partial E_y}{\partial x}-\frac{\partial E_x}{\partial y}=0
\end{equation}

At the differential scale, a variation of one component of the electric field under displacements in one direction implies a variation of another component of the same field under displacements in another direction. This is a point worth highlighting regarding the \textit{structure} of the electrostatic field.\\

\end{multicols}


\begin{multicols}{2}

We launch our series of invitations to discuss issues of
interest through questions related to Maxwell's equations for
static fields. \\

\textbf{Introductory question:} In the integral form of Gauss's
Law of Electricity ($\varepsilon_0 \oint \mathbf{E}\cdot d\mathbf{a}=q$), is $\mathbf{E}$ attributable to q? \\

\textbf{Topics for discussion:}

-Is the expression “the field associated to a charge” better
than the popular “the field produced by a charge? This
question has to do with causality. It is worth considering it in
the discussion of any Maxwell equation. The search for a
correct answer can lead us to Quantum Electrodynamics \cite{14}.

-What kind of new physical phenomena would happen if
equation (\ref{eqn:5}) was satisfied by an integrated electro-magnetic
field in a combined space-time?
\section{Time-varying fields. The laws of
Faraday and Ampere-Maxwell}

Faraday’s remarkable proposition of the concept of lines of force bridging the space between interacting charges, currents, and magnets established the starting point for the transcendental work done by Maxwell. Maxwell mathematized this idea and built a unified theory of electromagnetism. The concept of the electromagnetic field,
including the novel but essential displacement current (${\bf{J_D}}=\mu_0 \varepsilon_0 \frac{\partial \mathbf{E}}{\partial t}$), is central to Maxwell's theory. \\

The general laws of Electromagnetism, including time-dependent interactions, are formed by Gauss’s Laws (presented in Table \ref{table:1}, valid throughout the whole of
Electrodynamics), plus Faraday’s Law for induction and Ampere’s Law (generalized by Maxwell for $\mathbf{B}$ induced by a changing $\mathbf{E}$). Table \ref{table:2} presents the time dependent Maxwell’s Equations.

A century and a half of experiments, theory, teaching and applications (cellular phones, electric cars, synchrotrons...) of electromagnetism are based on the equations in Tables \ref{table:1} and \ref{table:2}. \\

\begin{figure} [H]
\includegraphics[width=\linewidth]{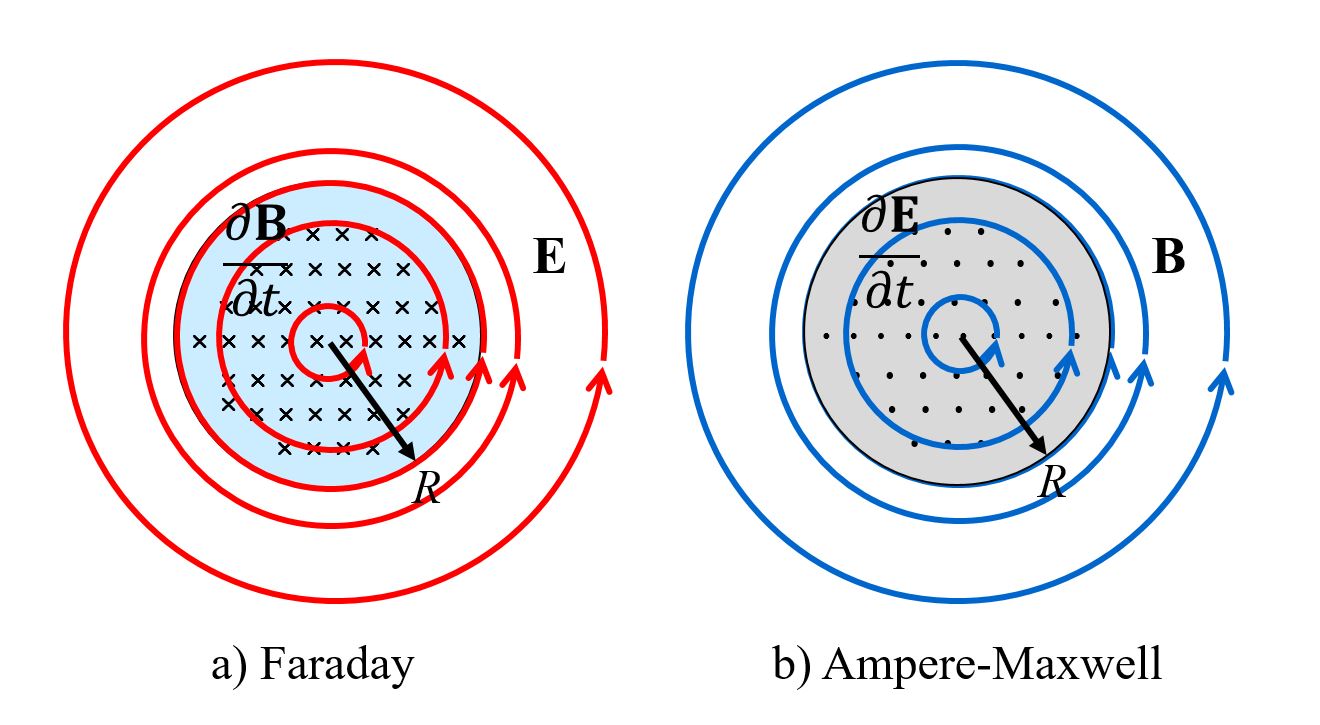}
\caption{Electromagnetic induction in configurations of
cylindrical symmetry.}
\label{fig:img2}
\end{figure}

Electric and magnetic fields ($\mathbf{E}$, $\mathbf{B}$) associated with time variations of ($\mathbf{B}$, $\mathbf{E}$) respectively, are commonly denoted “induced” fields. Figure \ref{fig:img2} represents induced fields corresponding to cylindrical symmetries.

The induced $\mathbf{E}$ and $\mathbf{B}$ both run circularly, counter clockwise, according to the respective associated (“inducing”) fields $\frac{\partial \mathbf{B}}{\partial t}$ and $\frac{\partial \mathbf{E}}{\partial t}$. Equations (\ref{eqn:6}) and (\ref{eqn:7}) represent the modules of these vectors as functions of the distances to the symmetry axes of each configuration.

\begin{equation}
\label{eqn:6}
    E(r)= \begin{Bmatrix}
\frac{r}{2}\frac{dB}{dt} & r < R \\
\frac{R^2}{2r}\frac{dB}{dt} & r> R
\end{Bmatrix}
\end{equation}

\begin{equation}
\label{eqn:7}
    B(r)= \begin{Bmatrix}
\frac{\mu_0\varepsilon_0 r}{2}\frac{dE}{dt} & r < R \\
\frac{\mu_0\varepsilon_0R^2}{2r}\frac{dE}{dt} & r> R
\end{Bmatrix}
\end{equation}

A couple of conceptual comments and a discussion topic
are worth introducing: \\

-As described by equation (\ref{eqn:6}), an electric field is found in the
region $r > R$, where $B = 0$. This field is linked with a changing
magnetic field in \textit{another} region of space ($r < R$).
Analogously, a varying electric field in $r < R$ induces a
magnetic field, even for $r > R$. The local relationship between
the components of vectors $\mathbf{E}$ and $\mathbf{B}$ shape the global structure
of the electromagnetic field.\\

-\textit{Electromagnetic induction} is the answer to the question,
related to equation (\ref{eqn:5}), and posed in the previous Section.\\

\textbf{Topic for discussion:}

We question the cause-effect relationship in induction phenomena. Take for example Faraday's Law. How to interpret the mathematical expression of this law? In its usual
form, as shown in Table \ref{table:2}, the general interpretation is that a time-changing magnetic field induces an electric field. If we write this equation in a way that reminds us of Newton's 2nd Law ($d\mathbf{p}/dt=\mathbf{F}$), namely $\partial \mathbf{B}/\partial t=-\nabla\times\mathbf{E}$, it is valid to interpret
that the temporal variation of the magnetic field is determined by certain local spatial variations of the electric field. This discussion is proposed by Hill \cite{15,16}, applying an approach consistent with Jefimenko's alternative formulation of electromagnetism \cite{17}.

\begin{table*}
\renewcommand{\arraystretch}{2}
\centering
\caption{Faraday and Ampere-Maxwell equations}
\label{table:2}
\vspace{0.3cm}
\begin{tabular}{p{5cm}p{5cm}p{5cm}}
\hline
Equation & Differential form & Integral form \\
 \hline
Faraday & $\nabla\times\mathbf{E}=-\frac{\partial\mathbf{B}}{\partial t}$ & $\oint\mathbf{E}\cdot d\mathbf{l}=-\frac{d\phi_B}{dt}$ \\
 \hline
Ampere-Maxwell & $\nabla\times\mathbf{B}= \mu_0 \mathbf{J}+\mu_0\varepsilon_0\frac{d\mathbf{E}}{dt}$ & $\oint\mathbf{B}\cdot d\mathbf{l}=\mu_0I+\mu_0\varepsilon_0\frac{d\phi_E}{dt}$  \\
 \hline
\end{tabular}
\end{table*} 

\section{Relativity of E and B}

The following is a fragment of the already cited EIS given by Einstein at Columbia University, New York (\cite{8} page 113): \\

\textit{“The special aim which I have constantly kept before me is logical unification in the field of physics. To start with, it disturbed me that electro dynamics should pick out one state of motion in preference to others, without any experimental justification for this preferential treatment. Thus arose the
special theory of relativity, which, moreover, welded together into comprehensible unities the electrical and magnetic fields, as well as mass and energy, or momentum and energy, as the case may be.”}\\

Also of interest is the renowned textbook “The Feynman Lectures on Physics”, which was published some forty years later. It presents, at undergraduate level, among several clarifying explanations, the relativistic relationship between electricity and magnetism \cite{18}. In Section 13-6, Prof.Feynman shows how the magnetic action of a current-carrying wire on a mobile charge may be predicted on the basis of charge invariance, Gauss’s law for electricity and Lorentz contraction. Today, a broad spectrum of levels and treatments for this subject can be found. The following
contributions typify from introductory \cite{19,20,21} to advanced \cite{22,23} available presentations.

\begin{figure*}[t]
\centering
\includegraphics[width=0.8\linewidth]{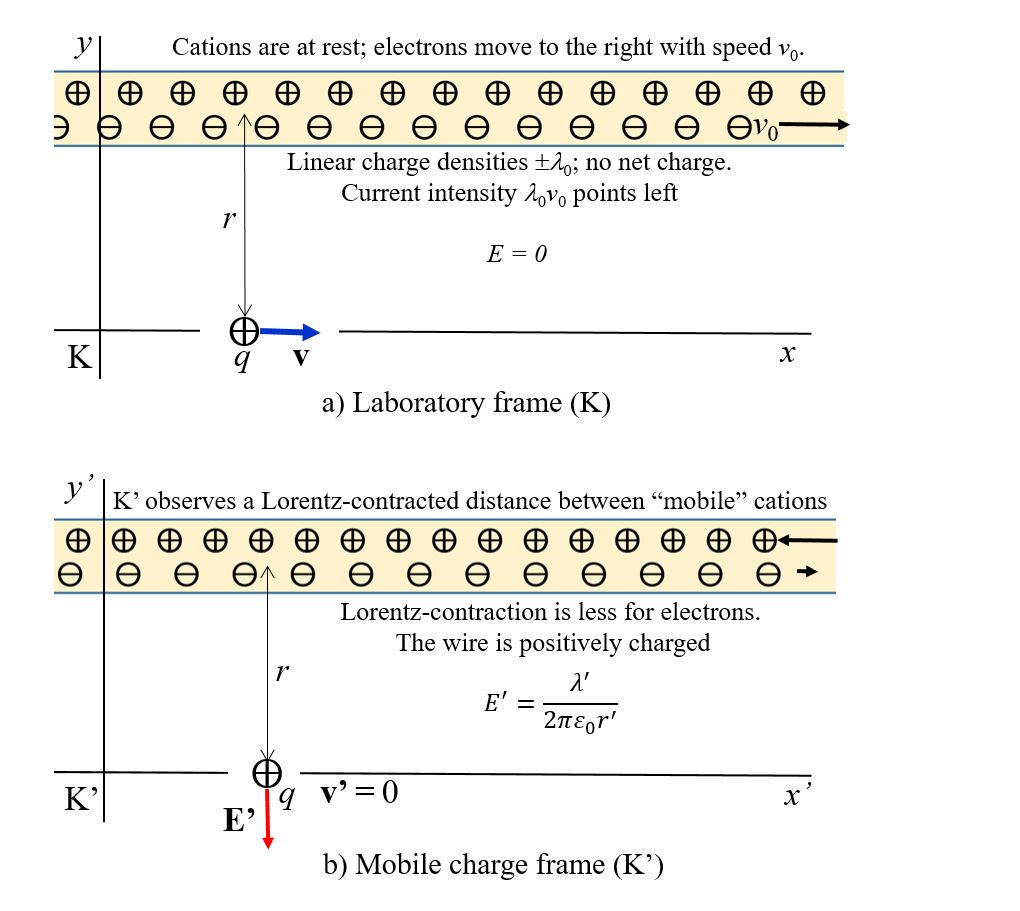}
\caption{Interaction of a moving charge with an electric current}
\label{fig:img3}
\end{figure*}

Below is a compact analysis of the topic, focusing on the interaction between an electric current and a moving charge. In Figure \ref{fig:img3} the electric current distribution of Figure \ref{fig:img1} is detailed in side view. The conductor contains positive (cations) and negative (electrons) charges. \\

Figure \ref{fig:img3}a shows the description from the "Laboratory" frame (K). The cations are at rest and the electrons are moving to the right at a speed $v_0$. Linear charge densities are equal to $\pm \lambda_0$. The conductor has no net charge but carries a current $I=\lambda_0v_0$ to the left (off the drawing in Figure \ref{fig:img1}b). The positive
charge \textit{q}, at a distance \textit{r} from the conductor, is moving to the right with a speed \textit{v}. Even though $\mathbf{E} = 0$, we shall now “discover” the existence of a force acting on \textit{q}, by purely electrical and relativistic arguments.

We call K’ the “Mobile charge” frame, as shown in Figure \ref{fig:img3}b. The tasks to be accomplished are: Find how K’ sees the charge distribution, find $\mathbf{E'}$ and also $\mathbf{F'} = q\mathbf{E'}$. Finally, transform $\mathbf{F'}$ to $\mathbf{F}$.\\

When we stand in the K’ system and look at the charges in this configuration, we can see a significant difference from what is observed at K. The conductor is no longer electrically neutral, it is charged. This happens as a combined result of the charge invariance and the Lorentz contraction. \\

For K’, the negative charges in the configuration move more slowly than for K, so the distance between them appears greater than for the observer in the laboratory. On the other hand, K’ sees positive charges moving, so it sees them closer together. Uniting these considerations of relative distance with that of relativistic invariance of the charge, it is concluded that, for the mobile charge (K’), the “conductor” has a positive charge. Performing the calculation of the net charge density observed from K’, it is found:

\begin{equation}
\label{eqn:8}
    \lambda ' =\frac{\gamma v\lambda_0 v_0}{c^2}=\frac{\gamma vI}{c^{2}}
\end{equation}
\noindent
where:

\begin{equation}
\label{eqn:9}
    \gamma=\frac{1}{\sqrt{1-(v^2/c^2)}}=\frac{1}{\sqrt{1-\beta^2}}; \beta = \frac{v}{c}
\end{equation}
\noindent
and \textit{c} is the speed of light in vacuum. \\

We already know how K’ sees the configuration: it sees it
as a positively charged infinite wire, with a linear density $\lambda'$ given by (\ref{eqn:8}). The field \textit{E’} in K’ is sought by applying Gauss’s Law, equation (3):

\begin{equation}
\label{eqn:10}
    E'=\frac{\lambda'}{2\pi\varepsilon_0r'}=\frac{Iv\gamma}{2\pi\varepsilon_0c^2r^\cdot}
\end{equation}

From here we obtain that \textit{q} (positive) will be repelled by
the current, with a force of module:

\begin{equation}
\label{eqn:11}
    F'=qE' = \frac{Iqv\gamma}{2\pi\varepsilon_0c^2r}
\end{equation}
\noindent
as measured by K’.

To find the force on \textit{q}, measured from K, the relativistic force transformation law is applied. The result is the following:

\begin{equation}
\label{eqn:12}
    F=\frac{1}{\gamma}F'=\frac{Iqv}{2\pi\epsilon_0c^2r}=\frac{\mu_0Iqv}{2\pi r}; c=\frac{1}{\sqrt{\mu_0\varepsilon_0}}
\end{equation}

Equation (\ref{eqn:12}) coincides with the one that would have been obtained by applying the magnetic force $\mathbf{F}=q\mathbf{v}\times\mathbf{B}$, with the vector $\mathbf{B}$ given by equation (\ref{eqn:4}). The curious thing about the case is that result (\ref{eqn:12}) was obtained \textit{without} using the idea of a magnetic field.

The following question is pertinent: Is $\mathbf{B}$ introduced for convenience, to avoid looking for a system with \textit{q} at rest? In other words, would the concept of a magnetic field be dispensable?

No. Einstein’s words quoted at the beginning of this section make this clear. A fundamental problem of electrodynamics is that of the structure of the field which
enables the interaction between charges. Thus, the accurate field must be one that correctly describes the interaction between charges, in any dynamic situation and from any inertial frame. Only the combination of $\mathbf{E}$ and $\mathbf{B}$ satisfies this condition.

In our particular case, observing from the laboratory system, there is $\mathbf{B}$, but $\mathbf{E} = 0$. In the mobile charge system, $\mathbf{E}$
and $\mathbf{B}$ exist (are different fron zero). The latter does not act
only because the charge is stationary.\\

\textbf{Topic for discussion:} \\

Some authors summarize what was analyzed here with the scheme [Electricity+ Relativity→ Magnetism] \cite{24,25}. On the other hand, \cite{26} demonstrates the mathematical validity of the opposite hypothesis [Magnetism + Relativity→ Electricity]. Who is right?\\

We present formulas (\ref{eqn:13}). They are the relativistic transformation equations of the electromagnetic fields between two inertial reference systems.

\begin{equation}
\label{eqn:13}
\begin{array}{l}
\left.
\begin{array}{l}
E'_x = E_x \\
E'_y = \gamma \left( E_y - v B_z \right) \\
E'_z = \gamma \left( E_z + v B_y \right) \\
B'_x = B_x \\
B'_y = \gamma \left( B_y + (v/c^2) E_z \right) \\
B'_z = \gamma \left( B_z - (v/c^2) E_y \right)
\end{array}
\right\}
\end{array}
\end{equation}

The equations (\ref{eqn:13}) have a high degree of generality, which is consistent with the field concept. If the components of ($\mathbf{E}$, $\mathbf{B}$) are known, at a point in space-time, measured with respect to an inertial reference system, these equations allow the components of ($\mathbf{E'}$, $\mathbf{B'}$) to be calculated at the same point of space-time, measured from another inertial system, which moves with speed $v$ with respect to the first. The application of (\ref{eqn:13}) does not require \textit{any} knowledge about the distant charges or currents that gave rise to these fields. This property‒the local character of the relativistic transformation laws‒is not inherent to $\mathbf{E}$ or $\mathbf{B}$ separately.

The approximate transformation equations for the case of
low velocities ($\beta=v/c\ll 1; \gamma \rightarrow 1$), are:

\begin{align}
\label{eqn:14}
    \mathbf{E'}&=\mathbf{E}+\mathbf{v}\times\mathbf{B} \\ 
    \mathbf{B'}&=\mathbf{B} \nonumber
\end{align}

The practical applicability of (\ref{eqn:14}) extends to the
approximate limit $\beta\sim 0.1$.\\

\textbf{Topic for discussion:} \\

It is worth comparing the interpretations of the transformation equations given in different texts \cite{23,11}. If it is considered that the relativistic transformation relations of a physical field must have a local character, is the electromagnetic field a physical field? And the electric one?

The integration as a concept of the electric and magnetic fields, and their interactions, is achieved compactly and elegantly via the relativistic tensor representation of the laws of electromagnetism. An introductory exposition of this formulation is presented next. We follow the lectures by Tong \cite{27} and Hughes \cite{28}

\section{Maxwell equations in tensor form}

The space of relativity is space-time, four-dimensional. The 4-vector $x^\eta$ denotes the position of a point in this space:

\begin{equation}
\label{eqn:15}
    x^\eta = (ct, x, y, z).
\end{equation}
\noindent
Greek indexes = 0, 1, 2, 3.

Therefore, the electromagnetic field, a unified entity, is
characterized by the electromagnetic field tensor:

\begin{equation}
\label{eqn:16}
    F^{\eta \xi}= \begin{pmatrix}
0 & -\nicefrac{E_x}{c} & -\nicefrac{E_y}{c} & -\nicefrac{E_z}{c}\\
\nicefrac{E_x}{c} & 0 & -B_z & B_y\\
\nicefrac{E_y}{c} & B_z & 0 & -B_x\\
\nicefrac{E_z}{c} & -B_y & B_x & 0
\end{pmatrix}
\end{equation}

The sources of the electromagnetic field are also described by a unified magnitude, the 4-vector current density $J^\eta$:

\begin{equation}
\label{eqn:17}
    J^\eta = (c\rho, J_x, J_y, J_z).
\end{equation}

A variant of $F^{\eta \xi}$, named its dual and denoted $\widetilde{F}^{\eta \xi}$ or $*F^{\eta \xi}$, is also functional. $\widetilde{F}^{\eta \xi}$ is obtained from $F^{\eta \xi}$ by means of the substitutions $E \rightarrow cB$ and $B \rightarrow -E/c$ :

\begin{equation}
\label{eqn:18}
    F^{\eta \xi}= \begin{pmatrix}
0 & -B_x & -B_y & -B_z \\
B_x & 0 & \nicefrac{E_z}{c} & -\nicefrac{E_y}{c} \\
B_y & -\nicefrac{E_z}{c} & 0 & \nicefrac{E_x}{c} \\
B_z & \nicefrac{E_y}{c} & -\nicefrac{E_x}{c} & 0 
\end{pmatrix}
\end{equation}

In the tensor formulation, the Maxwell equations are
expressed, in progressively compact notations, as follows:

\begin{align}
\label{eqn:19}
    \sum_\eta\frac{\partial F^{\eta\xi}}{\partial x^\eta}&= \frac{\partial F^{\eta\xi}}{\partial x^\eta}= \partial_\eta F^{\eta\xi} = \mu_0J^\xi \\
\label{eqn:20}
    \sum_\eta\frac{\partial \widetilde{F}^{\eta\xi}}{\partial x^\eta} &= \frac{\partial \widetilde{F}^{\eta\xi}}{\partial x^\eta} = \partial_\eta \widetilde{F}^{\eta\xi}= 0
\end{align}

Accordingly, Maxwell's equations in their usual form are obtained by expanding equations (\ref{eqn:19}) and (\ref{eqn:20}).

The Gauss Law for electricity is deduced from (\ref{eqn:19}). Calculating the case $\xi=0$:
 \begin{equation}
 \label{eqn:21}
     \frac{\partial E_x}{c \partial x} + \frac{\partial E_y}{c \partial y} + \frac{\partial E_z}{c \partial z} = \mu_0c\rho \rightarrow \nabla \cdot \mathbf{E} = \frac{\rho}{\varepsilon_0}.
 \end{equation}

The Ampere-Maxwell Law corresponds to the cases $\xi=1, 2, 3$. We expand $\xi =1$:

\begin{align}
\label{eqn:22}
    -\frac{\partial E_x}{c^2\partial t} + \frac{\partial B_z}{\partial y} &- \frac{\partial B_y}{\partial z} = \mu_0J_x \nonumber \\
    &\downarrow \nonumber \\  
    \nabla \times \mathbf{B} = \mu_0\mathbf{J}& + \mu_0\varepsilon_0\frac{\partial\mathbf{E}}{\partial t}.
\end{align}

The Gauss Law for magnetism and Faraday's Law are derived from the equation (\ref{eqn:20}).

\begin{align}
\label{eqn:23}
    &\xi = 0: \quad && \nonumber \\
    &\hspace{1cm} \frac{\partial B_x}{\partial x}+\frac{\partial B_y}{\partial y}+\frac{\partial B_z}{\partial z}=0 \rightarrow \nabla\cdot\mathbf{B}=0 &&
\end{align}

\begin{align}
\label{eqn:24}
    &\xi = 1: \quad && \nonumber \\
    &\hspace{0.5cm} -\frac{\partial B_x}{c\partial x}-\frac{\partial E_z}{c\partial y}+\frac{\partial E_y}{c\partial z}=0 \rightarrow \nabla\times\mathbf{E}=-\frac{\partial\mathbf{B}}{\partial t}. &&
\end{align}

\section{Conclusions}

This article proposes several tools to help improve students' understanding of some fundamental aspects of the laws of electromagnetism. Included are quotes from Einstein and Bohr on the topics analyzed, the raising of questions open to discussion and numerous citations to historical and current contributions of interest. The class discussion of the proposed questions will probably promote research and deeper learning. The citations (the majority with effective links) may form a guide to further explorations.\\

\noindent
We suggest that the following concepts should be highlighted
in the development of the electromagnetic course:\\

- Since Newton, a fundamental goal of physics has been to establish descriptions, at the infinitesimal level of detail, of the cause-effect relationships and the structure of the objects of the material world. The proper mathematical framework to
fulfill this purpose comprises differential equations.

- The electromagnetic theory forms a conceptual body consistent with Einstein’s special relativity. In this framework, the physical field has electromagnetic nature. Its structure, at the local level in space-time, is described by Maxwell’s equations. The infinitesimal variations of the electromagnetic field generate its structure on a global, macroscopic scale.\\

\noindent
A couple of cases, representative of the relevance of the local
character of the field properties, are:

- The information of what happens in the vicinity of a point in space-time characterize the interaction between the field components. This point is worth mentioning from the analysis of time-independent fields, but reveals itself in a colorful manner in the study of electromagnetic induction.

- The components of the field are transformed relativistically. The relativistic transformation law (considering both the electrical and magnetic components) has a local character. The interaction via electromagnetic field is correctly characterized from any inertial system.

- Maxwell’s equations acquire their most encompassing expression in the tensor formalism of Einstein’s relativity.

\section*{Acknowledgement}

Both authors are thankful to Ruth Chavira, the outstanding student who translated this article to latex code. We would like to also acknowledge the referee for the comments. We consider that the article has significantly improved its quality thanks to suggestions. Support from CONAHCYT, Project CF/2019 1085 is recognized as well.

\section*{Supplementary information}

The Appendix 1 consists of a pictorial tribute to Newton, Maxwell and Einstein for some of their fundamental contributions, expressed in mathematical language, highlighted in the present article. Appendix 2 shows the conversion of several electromagnetic magnitudes from Maxwell’s notation to current symbology.

\end{multicols}

\begin{center}
    
\subsection*{Appendix 1}

Newton, Maxwell, Einstein and a selection of fundamental equations of physics, associated with their contributions, commented in the present article.

\begin{figure*}[h]
\centering
\includegraphics[width=\linewidth]{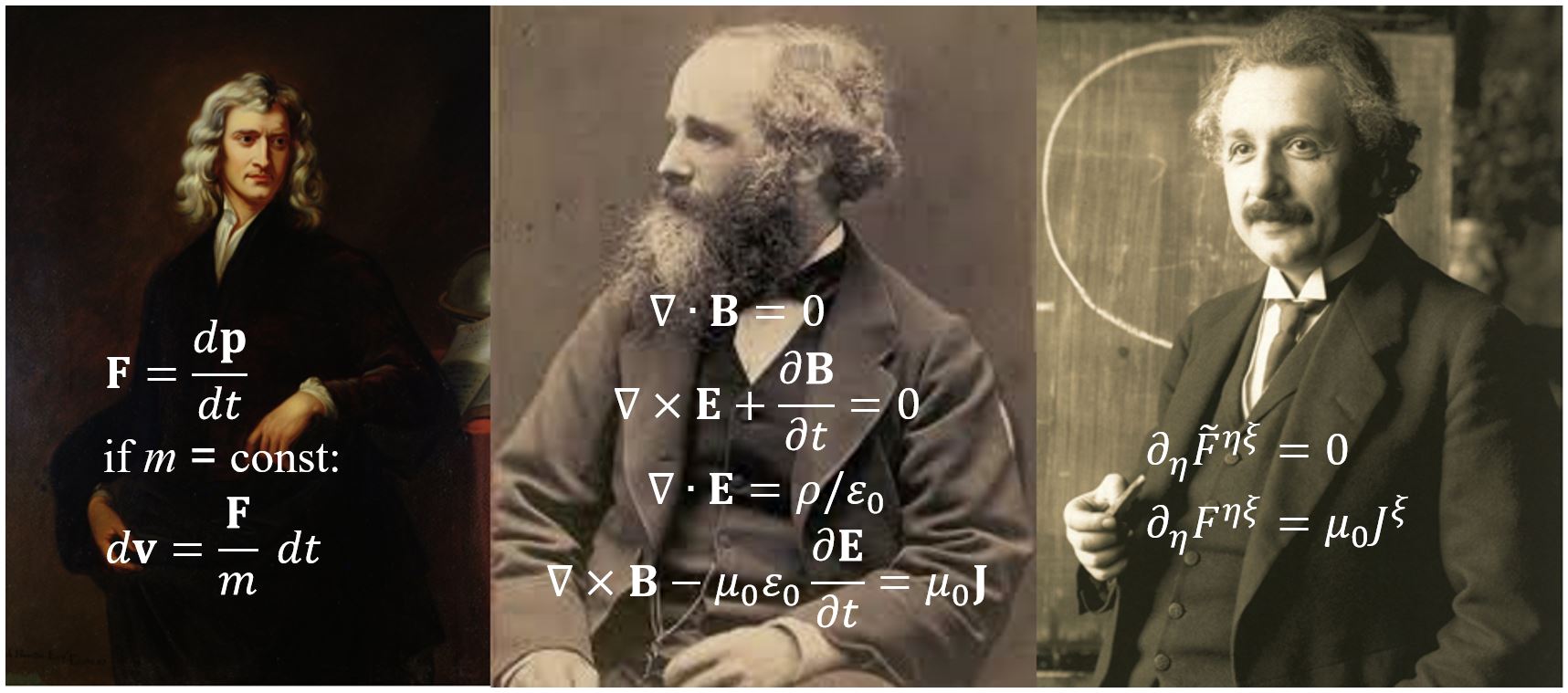}
\label{fig:imgA}
\end{figure*}
\newpage
\subsection*{Appendix 2}

\textbf{The original and present day forms of electromagnetic magnitudes}

\end{center}

The objective of the present Appendix is to facilitate the reading of transcendental segments in Maxwell’s original writings. In his landmark paper \cite{29}, Maxwell characterized fields through their components. As an example, we briefly
compare the original symbols in the (currently denoted) Ampere-Maxwell equation with their present day notation. Maxwell represented it by means of expressions of the type of the equations (A1):

\begin{equation} \tag{A1}
\label{eqn:A1}
\begin{array}{l}
\left.
\begin{array}{l}
\frac{d\gamma}{dy}-\frac{d\beta}{dz}=4\pi p' \\
\frac{d\alpha}{dz}-\frac{d\gamma}{dx}=4 \left(\pi q' \right) \\
\frac{d\beta}{dx}-\frac{d\alpha}{dy}=4 \left(\pi r' \right).
\end{array}
\right\}
\end{array}
\end{equation}

The magnitudes in (\ref{eqn:A1}), in Maxwell’s article, have the following meaning: \\
\newline
$\alpha, \beta, \gamma \rightarrow$ components of the intensity of the magnetic field
(nowadays $H_1$, $H_2$, $H_3$) 
\newline
$p'=p+df/dt.$
\newline
$p, q, r \rightarrow$ electric current density components (currently $j_1$, $j_2$, $j_3$)
\newline
$f, g, h \rightarrow$ components of electric displacement (today $D_1$, $D_2$,
$D_3$).\\

Table A1 shows the conversion to the current nomenclature (in the International System of Units) of a selection of quantities presented in the Maxwell’s originals. The contemporary notation for Maxwell theoretical ideas is due to Heaviside \cite{30,31}. The way Faraday and Ampere-Maxwell Laws appear in Heaviside’s book (p. 130) is as follows:

\begin{equation} \tag{A2}
\label{eqn:A2}
    -curl \mathbf{E}=\mu{\bf\dot{H}}
\end{equation}
\begin{equation} \tag{A3}
\label{eqn:A3}
    curl \mathbf{H}=k\mathbf{E}+c{\bf\dot{E}}
\end{equation}

In the present-day symbology, $\mu$ is the magnetic permeability, $k$ is the electrical conductivity and $c$ can be related to the product of the permittivity and the permeability.\\

Table A1 shows the equivalence between several symbols used by Maxwell and the ones of current usage.
\begin{table*}
\renewcommand{\arraystretch}{2}
\centering
Table A1: Electromagnetic magnitudes in Maxwell’s notation and in current symbology
\vspace{5pt}
\begin{tabular}{p{4cm}p{4cm}|p{4cm}p{4cm}}
\hline
\multicolumn{2}{c|}{\textbf{Maxwell works}} & \multicolumn{2}{c}{\textbf{Current nomenclature (SI)}}  \\
 \hline
\textbf{Description}& \textbf{Symbols} & \textbf{Description}  & \textbf{Symbols} \\
\hline
Quantity of Free Electricity & $e$ & Electric Charge Density & $\rho=dq/dV$ \\
\hline
Electric Potential & $\psi$ & Electric Potential & $\phi, \varphi$ \\
\hline
Electromotive Force & $P, Q, R$ & Electric field & $\mathbf{E}=[E_1, E_2, E_3]$ \\
\hline
Electric Displacement & $f, g, h$ & Electric Displacement & $\mathbf{D}=[D_1, D_2, D_3]\newline=\varepsilon_0\mathbf{E+\mathbf{P}}$ \\
\hline
Electric Elasticity & $k \rightarrow P=kf, Q=kg, \newline R= kh$ & Electric Permittivity & $\mathbf{D}=\varepsilon\mathbf{E}$ (linear isotropic) [$\varepsilon=l/k$(Maxwell)]  \\
\hline
Electric Current & $p, q, r$ & Electric Current
Density & $\mathbf{J}=[j_1, j_2, j_3]$ \\
\hline
Total Electric Current & $p'=p+df/dt$, similar for $q', r'$  & Total Electric Current
Density & $\mathbf{J}_T=\mathbf{J}+d\mathbf{D}/dt$ \\
\hline
Magnetic Intensity = Magnetic Force & $\alpha, \beta, \gamma$ & Magnetic Intensity & $\mathbf{H}=[H_1, H_2, H_3]$ \\
\hline
Magnetic Induction & $\mu\alpha, \mu\beta, \mu\gamma$ & Magnetic Induction & $\mathbf{B}=[B_1, B_2, B_3] \newline = \mu_0(\mathbf{H}+\mathbf{M})$ \\
\hline
Coefficient of Magnetic Induction & $\mu$ & Magnetic permeability & $\mathbf{B}=\mu\mathbf{H}$ (linear-isotropic) \\
\hline
Equations of Currents & $\frac{d\gamma}{dy}-\frac{d\beta}{dz}=4\pi p'$, similar for $q'$, $r'$  & Ampere-Maxwell Equation & $\nabla\times\mathbf{H}=J_T=\mathbf{J}+\frac{d\mathbf{D}}{dt}$ \\
\hline
Electric Resistance & $P=-\rho p, Q =-\rho q, \newline R=-\rho r$  & Electric Resistivity & $\mathbf{E}=\rho\mathbf{J}$ \\
\hline
Equation of Free Electricity & $e+\frac{df}{dx}+\frac{dg}{dy}+\frac{dh}{dz}=0$ & Gauss Law for Electricity & $\nabla\cdot\mathbf{D}=\rho$ \\
\hline
Equation of Continuity & $\frac{de}{dt}+\frac{dp}{dx}+\frac{dq}{dy}+\frac{dr}{dz}=0$ & Equation of Continuity & $\frac{d\rho}{dt}+ \nabla\cdot\mathbf{J}=0$ \\
\hline
\end{tabular}
\end{table*} 

\newpage
\begin{multicols}{2}

\end{multicols}

\section{References}

    
\begin{thebibliography}{31}
\bibitem{1} L. Bollen, P. Van Kampen, C. Baily, M. Kelly, and M. De Cock, Physical Review Physics Education Research
\textbf{13}, 020109 (2017).  \url{https://doi.org/10.1103/PhysRevPhysEducRes.13.020109}
\bibitem{2} B. Buonaura, and G. Giuliani, DOI
\url{http://dx.doi.org/10.1393/gdf/i2024-10535-8}, (2024).
\url{http://dx.doi.org/10.1393/gdf/i2024-10535-8}
\bibitem{3} J. Martí, Obras completas \textbf{8}, 281-284 (1883).
\url{https://martianoscuba.wordpress.com/wp-content/uploads/2022/06/vol08.pdf}
\bibitem{4} F. Richtmyer, American Journal of Physics \textbf{1}, 1-5
(1933).
\url{https://doi.org/10.1119/1.1992814}
\bibitem{5} V. F. Weisskopf, Physics Today \textbf{29}, 23-29 (1976). 
\url{https://doi.org/10.1063/1.3023516}
\bibitem{6} C. Hidalgo, EPS Grand Challenges Physics for Society in the Horizon 2050, (IOP Publishing, Bristol,
UK, 2024), pp. 831.
\url{https://doi.org/10.1088/978-0-7503-6342-6}
\bibitem{7} M. F. Taşar, and P. R. Heron, The International Handbook of Physics Education Research: Special Topics, (AIP Publishing Books, 2023), pp. 6.1-26.24. 
\url{https://doi.org/10.1063/9780735425514}
\bibitem{8} A. Einstein, Essays in science, (Philosophical
Library, New York, 1934), pp. 28-45, 112-114.
\url{https://archive.org/details/essaysinscience0000eins}
\bibitem{9} M. Bunge, Critical approaches to science and philosophy, (Routledge, 2018), pp. 234-243.
\url{https://www.routledge.com/Critical-Approaches-to-Science-and-Philosophy/Bunge/p/book/9780765804273}
\bibitem{10} N. Bohr, Dialectica \textbf{2}, 312-319 (1948). 
\url{https://doi.org/10.1111/j.1746-8361.1948.tb00703.x}
\bibitem{11} E. M. Purcell, and D. J. Morin, Electricity and
magnetism, (Cambridge university press, Cambridge, 2013), pp. 98-99, 242-246, 259-267, 306-314.
\url{https://archive.org/details/ElectricityAndMagnetismPurcell_2013}
\bibitem{12} B. Crowell, Fields and Circuits, (Light and Matter, Fullerton, 2021), pp. 463.
\url{https://www.lightandmatter.com/fac/}
\bibitem{13} L. Fuentes-Cobas, J. Matutes-Aquino, and M. Fuentes-Montero, Handbook of magnetic materials \textbf{19}, 129-229 (2011).
\url{https://doi.org/10.1016/B978-0-444-53780-5.00003-X}
\bibitem{14} R. P. Feynman, QED: The strange theory of light and matter, (Princeton University Press, Princeton,
1985), pp. 158. 
\url{https://www.torrossa.com/en/resources/an/5575989}
\bibitem{15} S. E. Hill, The Physics Teacher \textbf{48}, 410-412 (2010).
\url{https://doi.org/10.1119/1.3479724}
\bibitem{16} S. E. Hill, The Physics Teacher \textbf{49}, 343-345 (2011).
\url{https://doi.org/10.1119/1.3628256}
\bibitem{17} O. D. Jefimenko, European journal of physics \textbf{25}, 287 (2004).
\url{https://doi.org/10.1088/0143-0807/25/2/015}
\bibitem{18} R. P. Feynman, The Feynman lectures on physics, (Addison-Wesley, Michigan, 1963), pp. 13.6-13.10.
\url{https://www.feynmanlectures.caltech.edu/II_13.html}
\url{https://www.feynmanlectures.caltech.edu/Notes.html}
\bibitem{19} D. Halliday, R. Resnick, and K. S. Krane, Physics, Volume 2, (John Wiley \& Sons, México, 2010), pp. 764- 765.
\url{https://archive.org/details/fisica-vol-2-halliday-resnick-and-kran/page/n209/mode/2up}
\bibitem{20} F. Kamphorst, M. Vollebregt, E. Savelsbergh, and
W. van Joolingen, Science \& Education \textbf{32}, 57-100 (2023).
\url{https://doi.org/10.1007/s11191-021-00283-2}
\bibitem{21} D. A. Muller, Magnets and Relativity, 2019.
\url{https://www.youtube.com/watch?v=1TKSfAkWWN0&t=174s&ab_channel=Veritasium}
\bibitem{22} D. Dugdale, Essentials of electromagnetism, (Macmillan, London, 1997), pp. 305-326.
\url{https://archive.org/details/essentialsofelec0000dugd}
\bibitem{23} D. J. Griffiths, Introduction to electrodynamics,
(Cambridge University Press, Cambridge, 2024), pp.554-572.
\url{https://doi.org/10.1017/9781009397735}
\bibitem{24} H. de Vries, The simplest, and the full derivation of Magnetism as a Relativistic side effect of ElectroStatics, 2008.
\url{http://chip-architect.org/physics/Magnetism_from_ElectroStatics_and_SR.pdf}
\bibitem{25} J. Houlihan, American Journal of Physics \textbf{90}, 248-248 (2022).
\url{https://doi.org/10.1119/5.0088991}
\bibitem{26} O. D. Jefimenko, European Journal of Physics \textbf{17},
180 (1996).
\url{https://doi.org/10.1088/0143-0807/17/4/006}
\bibitem{27} D. Tong, Lectures on Electromagnetism, (University of Cambridge, Cambridge, 2024), pp. 95-115.
\url{https://www.damtp.cam.ac.uk/user/tong/em.html}
\bibitem{28} S. A. Hughes, A covariant Formulation of electromagnetics, MIT, Boston, 2021.
\url{https://web.mit.edu/sahughes/www/8.033/lec12.pdf}
\bibitem{29} J. C. Maxwell, Philosophical transactions of the Royal Society of London \textbf{155}, 459-512 (1865).
\url{https://archive.org/details/dynamicaltheoryo00maxw}
\bibitem{30} H. Chaparro Hernández, and E. A. Meza Lozano, Aportes de Oliver Heaviside a la teoria electromagnetica de Maxwell y a su enseñanza, (Universidad Pedagógica Nacional, Bogotá, 2015), pp. 68.
\url{http://repository.pedagogica.edu.co/handle/20.500.12209/2125}
\bibitem{31} O. Heaviside, Electromagnetic theory, (The Electrician, London, 1893), pp. 130.
\url{https://archive.org/details/electromagnetict02heavrich/electromagnetict02heavrich/}

\end{thebibliography}
\end{document}